\documentclass[pra,showpacs,twocolumn,amsmath,amssymb,floatfix,superscriptaddress]{revtex4-1}

\usepackage{graphicx}
\usepackage{pstricks,pst-node,pst-text,pst-3d}
\usepackage{color}
\graphicspath{{Figures/}}

\newcommand{\pdagger}{{\phantom{\dagger}}}

\newcommand{\reff}[1]{Fig.\ \ref{fig:#1}}

\newcommand{\myparagraph}[1]{{\it #1} -- }

\newcommand{\neel}{N\'{e}el}

\newcommand{\TND}{T_{\text{N}}^{\text{DMFT}}}

\newcommand{\NNcorr}{\langle\hat{\boldsymbol{\sigma}}_i\cdot\hat{\boldsymbol{\sigma}}_j\rangle}

\newcommand{\nop}[1]{}

\begin{document}

	\title{Universal probes for antiferromagnetic correlations and entropy\\
	       in cold fermions on optical lattices}

\author{E.~V.~Gorelik}
\affiliation{Institute of Physics, Johannes Gutenberg University, Mainz, Germany}
\author{D.~Rost}
\affiliation{Institute of Physics, Johannes Gutenberg University, Mainz, Germany}
\author{T.~Paiva}
\affiliation{Instituto de Fisica, Universidade Federal do Rio de Janeiro, Brazil}
\author{R.~Scalettar}
\affiliation{Department of Physics, UC Davis, USA}
\author{A.~Kl\"umper}
\affiliation{University of Wuppertal, Wuppertal, Germany}
\author{N.~Bl\"umer}
\affiliation{Institute of Physics, Johannes Gutenberg University, Mainz, Germany}

\date{\today}
  \begin{abstract}
	We determine antiferromagnetic (AF) signatures in the half-filled
Hubbard model at strong coupling on a cubic lattice and in lower
dimensions. Upon cooling, the transition from the charge-excitation
regime to the AF Heisenberg regime is signaled by a universal minimum of
the double occupancy at entropy $s\equiv S/(Nk_{\text{B}}) = s^* \approx \ln(2)$ per
particle and a linear increase of the next-nearest neighbor (NNN) spin
correlation function for $s<s^*$. 
This crossover, driven by a gain in kinetic
exchange energy, appears as the essential AF physics relevant for
current cold-atom experiments. 
The onset of long-range AF order (at low $s$ on cubic lattices)
is hardly visible in nearest-neighbor spin correlations versus $s$, but
could be detected in spin correlations at or beyond NNN distances.
\end{abstract}
  \pacs{67.85.-d, 03.75.Ss, 71.10.Fd, 75.10.-b}
  \maketitle

%%%%%%%%%%%%%%%%%%%%%%%%%%%%%%%%%%%%%%%%%%%%%%%%%%%%%%%%%%%%%%%%%%%%%%%%
% Introduction
%%%%%%%%%%%%%%%%%%%%%%%%%%%%%%%%%%%%%%%%%%%%%%%%%%%%%%%%%%%%%%%%%%%%%%%%

Materials with strong electronic correlations are, due to their
increasing technological importance, a prime subject of current research
\cite{Tokura03,Dagotto05}.  Theoretical investigations of corresponding
Hubbard-type models have shed light on many strong-coupling phenomena,
including metal-insulator transitions, non-Fermi-liquid behavior, and
various types of magnetic and orbital order \cite{AnisimovBook2010}.
However, important questions remain open, most notably regarding
high-temperature superconductivity, for which so far no mechanism could
be established. 
Recently, a novel class of correlated Fermi systems,
namely ultracold fermionic atoms (such as $^{40}$K and $^6$Li) on
optical lattices, has opened a new promising direction of research: Cold
atoms are predicted to serve as {\it quantum simulators} for the Hubbard
type solid-state Hamiltonians of interest
\cite{Hofstetter_PRL02,Zoller05,Esslinger10}.

Indeed, the Mott metal-insulator transition has recently been
demonstrated in two-flavor mixtures of $^{40}$K on cubic optical
lattices by experimental observation and quantitative theoretical
analysis of signatures in the compressibility \cite{Bloch08_ferm} and in
the double occupancy \cite{Esslinger08_Nature}. This success established
that the single-band Hubbard model 
\begin{equation}\label{eq:Hubbard}
  \hat{H} =\! -t \sum_{\langle ij\rangle ,\sigma}  \hat{c}^{\dag}_{i\sigma}
\hat{c}^\pdagger_{j\sigma} 
+  U \sum_{i} \hat{n}_{i\uparrow} \hat{n}_{i\downarrow}
\end{equation}
(with hopping amplitude $t$, on-site interaction $U$, and
$\hat{n}_{i\sigma}=\hat{c}^{\dag}_{i\sigma}\hat{c}^\pdagger_{i\sigma}$)
can be realized to a reasonable accuracy using ultracold fermions in the
interesting interaction range, which certainly supports the hopes of
accessing also less understood Hubbard physics in similar ways.

However, attempts to realize and detect {\it quantum magnetism}
in cold lattice fermions have proven extremely challenging. 
In fact, it is even difficult to verify specific 
signatures of antiferromagnetic (AF) correlations which are  
believed to play an important role in high-temperature superconductivity.
This type of physics clearly has to be under control
before cold fermions on optical lattices can 
really play a useful role as {\it quantum simulators} 
of materials with strong electronic correlations. 
The failure to detect AF signals has primarily
been attributed to cooling issues \cite{Greif10,McKay11}. Indeed, the
coldest experiments for repulsive fermions on optical lattices 
have thus far reached central entropies per particle of
$s \equiv S/(Nk_B) \approx \ln(2)\approx 0.7$ \cite{Joerdens10,fn:entropy1} while
AF long-range order (LRO) on an isotropic cubic lattice is expected only for
entropies $s<s_{\text{N}}\approx 0.4$ \cite{Wessel10,Greif10,Fuchs11}.

An important feature of cold-atom systems is their inhomogeneity, induced by the trapping potential.
On the one hand, this is beneficial for quantum magnetism, since entropy is effectively pushed out of a half-filled core; this aspect is a major theme of current research \cite{Joerdens10,Fuchs11,Chiesa11,Paiva11,Khatami11}. 
On the other hand, any AF region will necessarily be of limited spatial extent \cite{fn:length}, so that the thermodynamic concept of LRO is not fully applicable. In addition, as we will show, the nearest-neighbor (NN) spin correlation function, which is currently addressed (using modulation spectroscopy \cite{Kollath06,Greif10} or superlattices \cite{Trotzky10}) in AF related experiments, is hardly sensitive to the onset of LRO even in the thermodynamic limit.

In this situation, one may ask the following: 
(i) Is there a threshold distance beyond which spin correlations have 
``long-range characteristics'' and (ii) can we define ``finite-range antiferromagnetism'' as a unique scenario with universal properties, appearing only in a certain entropy range? The answer to both questions is ``yes'': The essential AF correlation physics emerges already at entropies $s\lesssim \ln(2)$, i.e. in reach of current cooling techniques. Since, in addition, the threshold distance turns out to be rather small (but larger than one lattice spacing), our quantitative predictions should enable experimentalists to verify specific AF signatures with current system sizes, i.e., to get the long-sought grip on {\it quantum magnetism}.

%%%%%%%%%%%%%%%%%%% Overview %%%%%%%%%%%%%%%%%%%

In the following, we discuss first an enhancement of the double
occupancy $D$ (i.e. also of the interaction energy $E_{\text{int}}\equiv
D\, U$) at low temperatures $T$ which has previously been proposed as an
AF signature on the basis of dynamical mean-field theory (DMFT)
\cite{Gorelik_PRL10}. DMFT results for a half-filled cubic lattice at
strong coupling $U/t=15$ are compared with direct determinantal quantum
Monte Carlo (DQMC) \cite{Blankenbecler81,Paiva10} simulations. The
comparison is extended to the full dimensional range based on DQMC and
Bethe ansatz (BA) \cite{BA} data in dimensions $d=2$ and $d=1$,
respectively.  High-precision estimates of the entropy $s(T)$ allow us
to switch to the experimentally relevant entropy representation.  An
asymptotic collapse of the curves $D(s)$ is observed with respect to
dimensionality, with universal minima at $s^* \approx \ln(2)$, and no
significant features at $s_{\text{N}}$ in the cubic case. Additional
specific signatures of finite-range AF order are found in the kinetic
energy and in spin correlation functions, with different degrees of
universality. Finally, the perspectives for detecting LRO are discussed
using stochastic series expansion (SSE) results for the Heisenberg
model.

%%%%%%%%%%%%%%%%%%%%%%%%%%%%%%%%%%%%%%%%%%%%%%%%%%%%%%%%%%%%%%%%%%%%%%%%%%%%%%%%%
% AF signatures in the double occupancy
%%%%%%%%%%%%%%%%%%%%%%%%%%%%%%%%%%%%%%%%%%%%%%%%%%%%%%%%%%%%%%%%%%%%%%%%%%%%%%%%%

\myparagraph{AF signatures in the double occupancy.} 
According to DMFT, the low-$T$ formation of an AF core in a fermionic
cloud on an optical lattice (with central half filling, $n=1$) is
signaled, at strong coupling, by a distinct enhancement of $D$ in the
same region \cite{Gorelik_PRL10}.  As a function of temperature, DMFT
predicts nearly flat curves $D(T)$ in the range $T\gtrsim\TND$, i.e.
above its estimate of the \neel\ temperature, and a sharp increase
below, with a kink and absolute minimum at $\TND$.  This is clearly
seen, for $U/t=15$, in \reff{3d_SvsT}(a) (circles).  The absolute low-$T$
increase of $D$ is largest for $U/t\approx 12$; it should be detectable,
according to real-space DMFT, even in experiments integrating over the
inhomogeneous cloud \cite{Gorelik_PRL10}.

Not all aspects of this DMFT scenario are, however, realistic: After
all, DMFT is exact only in the limit of infinite coordination number
$Z\to\infty$ (with $Z=2d$ for hypercubic lattices) and overestimates the
\neel\ temperature by up to $30\%$ in the simple cubic case
\cite{Kent05,Staudt00}. Thus, the sharp kink in $D(T)$ seen in
\reff{3d_SvsT}(a) at $\TND\approx 0.4t$ cannot be physical. One might
expect a shift of the DMFT results toward {\it lower} temperatures, as
well as some broadening in the cubic case and more radical changes (at
least) for $d\le 2$; only at high temperatures does the accuracy of DMFT
estimates for $D$ follow already from series expansions (in $d=3$)
\cite{De_Leo_PRA_2011}. 

%%%%%%%%%%%%%%%%%%%%%%%%%%%%%%%%%%%%%%%%%%%%%%%%%%%%%%%%%%%%%%%%%%%%%%%%%%%%%%%%%
\begin{figure}[t]
\includegraphics[width=\columnwidth]{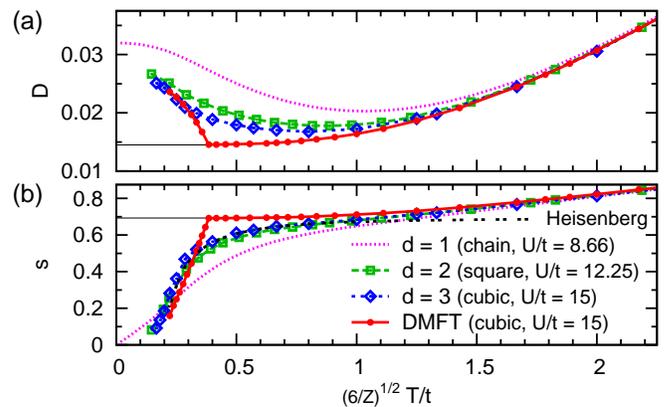}
\caption{(Color online)
Hypercubic lattice ($1\le d\le 3$) at strong coupling $U/(\sqrt{Z}t)\approx 6$:
(a) Double occupancy $D(T)$ as estimated from  DMFT ($d=3$, circles), 
DQMC ($d=3$: diamonds; $d=2$: squares), and BA ($d=1$, dotted line). 
(b) Corresponding estimates of  entropy per particle $s=S/(N k_\text{B})$. 
%All interactions correspond approximately to the ground state Mott transition at $U/(\sqrt{Z}t)\approx 6$. 
}\label{fig:3d_SvsT}
\end{figure}
%%%%%%%%%%%%%%%%%%%%%%%%%%%%%%%%%%%%%%%%%%%%%%%%%%%%%%%%%%%%%%%%%%%%%%%%%%%%%%%%%
\myparagraph{Impact of dimensionality.}
Indeed, DQMC estimates of $D(T)$ [diamonds in \reff{3d_SvsT}(a)] agree
with DMFT for $T/t\gtrsim 1$ within error bars, which are smaller than
symbol sizes \cite{fn:errors}.  Surprisingly good agreement is observed
also at $T/t\!\lesssim\! 0.3$. As expected \cite{Gorelik_PRL10}, the
DMFT kink is smeared out in the DQMC data toward a broad minimum.  At
suitably rescaled \cite{fn:scale} interactions, DQMC data for $d=2$
(squares) show remarkably similar behavior.  Only the case $d=1$ (dotted
line) deviates more drastically at intermediate and low $T$. Note that
the position of the minimum in $D(T)$ shifts {\it upward} with
decreasing $d$, i.e., opposite to the naive expectation.

As optical lattice and interactions are switched on for the ultracold
atoms in a nearly adiabatic process \cite{Joerdens10}, the entropy $s$
(and not $T$) is the experimentally relevant control parameter.
\reff{3d_SvsT}b shows numerically exact data for $s(T)$, obtained
directly for $d=1$ and via the thermodynamic relation
%\[
$S(\beta)=\ln(4) + \beta E(\beta) - \int_{0}^\beta d\beta' E(\beta')$
%\]
[with $\beta=1/(k_B T)$ and energy $E$] for $d=2,3$ and
DMFT. Again, the agreement between $d=2$ and $d=3$ is striking; the
latter results converge to the Heisenberg limit for $T\lesssim 0.8t$.
Remarkably, the proper DMFT solution (circles) is close to the DQMC
result for the cubic lattice (diamonds) at $T\lesssim 0.3t$; only the
metastable nonmagnetic DMFT solution (thin solid lines), considered in
previous studies \cite{De_Leo_PRA_2011}, remains far off.

%%%%%%%%%%%%%%%%%%%%%%%%%%%%%%%%%%%%%%%%%%%%%%%%%%%%%%%%%%%%%%%%%%%%%%%%%%%%%%%%%

\begin{figure}[t]
\includegraphics[width=\columnwidth]{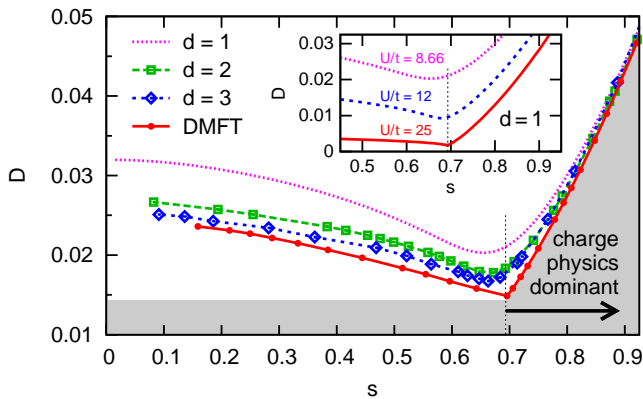}
\caption{(Color online)
Hypercubic lattice at strong coupling: 
Double occupancy vs entropy. In all cases, 
the minimum of the double occupancy corresponds to $s^* \approx \ln(2)$ (dotted line). 
The shaded area indicates the nonmagnetic contribution to $D$. 
Inset: $D(s)$ in $d=1$ for various interactions.
}\label{fig:dim_DvsS}
\end{figure}

Figure \ref{fig:dim_DvsS}, obtained by combining the data of both panels of
\reff{3d_SvsT}, conveys our first central message: As a function of
entropy, the double occupancy is surprisingly universal at strong
coupling, with a minimum at $s^* \approx \ln(2)$ in all dimensions and
generally similar shapes.  At constant rescaled interaction $U$, the
curvature around the minimum increases with increasing dimensionality
until it becomes sharp in the DMFT limit, where it corresponds to the
\neel\ transition. As seen in the inset (for $d=1$), the minimum becomes
also sharp and approaches $s^*$ at constant dimensionality in the
strong-coupling limit $U\to\infty$.  Evidently, the minimum in $D(s)$
separates two regimes with quite different physical properties: (i) the
regime  $s>s^*$, which smoothly approaches the Hartree limit (with
$D=\langle n_\uparrow\rangle\langle n_\downarrow\rangle=0.25$) for
$T\to\infty$, and (ii) a low-temperature regime, with no discernible
sub-structure. It is clear that the latter regime must be characterized
by spin coherence, since $s< \ln(2)$ can occur in a two-flavor system
at $n=1$ only by the development of (possibly short ranged) magnetic
correlations. 

%%%%%%%%%%%%%%%%%%%%%%%%%%%%%%%%%%%%%%%%%%%%%%%%%%%%%%%%%%%%%%%%%%%%%%%%%%%%%%%%%
%%%%%%%%%%%%%%%%%%%%%%%%%%%%%%%%%%%%%%%%%%%%%%%%%%%%%%%%%%%%%%%%%%%%%%%%%%%%%%%%%
In fact, any positive deviation of $D(s)$ from the nonmagnetic
background (shaded in \reff{dim_DvsS}) should be linked to AF
correlations, generalizing Takahashi's ground state expression
\cite{Takahashi77}
\begin{equation}\label{eq:Takahashi}
D_{0}=\frac{Z t^2}{2U^2}\left(1-\NNcorr_0^{\text{Heis}}\right)+\mathcal{O}\left(\frac{t^4}{U^4}\right).
\end{equation}
Here, $\NNcorr_0^{\text{Heis}}$ is the nearest-neighbor correlation in the quantum
Heisenberg model at $T=0$ (for Pauli matrices
$\hat{\boldsymbol{\sigma}}$), which is stronger in lower $d$: 
\[
\NNcorr_0^{\text{Heis}} = \left\{\!\begin{array}{llll}
-1.77 &\text{($d=1$) \cite{Takahashi77}},\, & -1.20 & \text{($d=3$) \cite{Oitmaa94}},\\ 
-1.34 &\text{($d=2$) \cite{Weihong91Sandvik97}}, & -1.00 & \text{($d=\infty$)\,,}
\end{array}\right.
\]
consistent with our finite-$T$ results.
Thus, irrespective of the measurement technique, signatures of AF
correlations may be easier to detect experimentally, at fixed $s$, for
lower (effective) dimensionality. Conversely, a tuning of the hopping
amplitude in $z$ direction could help to discriminate magnetic effects
from those of charge excitations; similar ideas involving frustration
will be explored in a separate publication \cite{Frustration}.

%%%%%%%%%%%%%%%%%%%%%%%%%%%%%%%%%%%%%%%%%%%%%%%%%%%%%%%%%%%%%%%%%%%%%%%%%%%%%%%%%
% AF signatures in other observables at strong coupling
%%%%%%%%%%%%%%%%%%%%%%%%%%%%%%%%%%%%%%%%%%%%%%%%%%%%%%%%%%%%%%%%%%%%%%%%%%%%%%%%%

\myparagraph{Energetics and spin correlations.} 
Up to a numerical prefactor (of $15/\sqrt{6}$), the curves in
\reff{dim_DvsS} represent the rescaled interaction energy
$E_{\text{int}}/(\sqrt{Z}t)$. Complementary AF signatures appear in the
kinetic energy, shown in \reff{dim_ENNvsS}(a). In particular,
$E_{\text{kin}}(s)$ exhibits a maximum at $s\approx s^*$ for $d\ge
3$ [while thermodynamic consistency ensures a monotonous total energy 
$E(s)=E_{\text{int}}(s)+E_{\text{kin}}(s)$]. The
associated redistribution of quasi momentum at $s\lesssim s^*$ might
be a worthwhile experimental target.
%%%%%%%%%%%%%%%%%%%%%%%%%%%%%%%%%%%%%%%%%%%%%%%%%%%%%%%%%%%%%%%%%%%%%%%%%%%%%%%%%
\begin{figure}[t]
\includegraphics[width=\columnwidth]{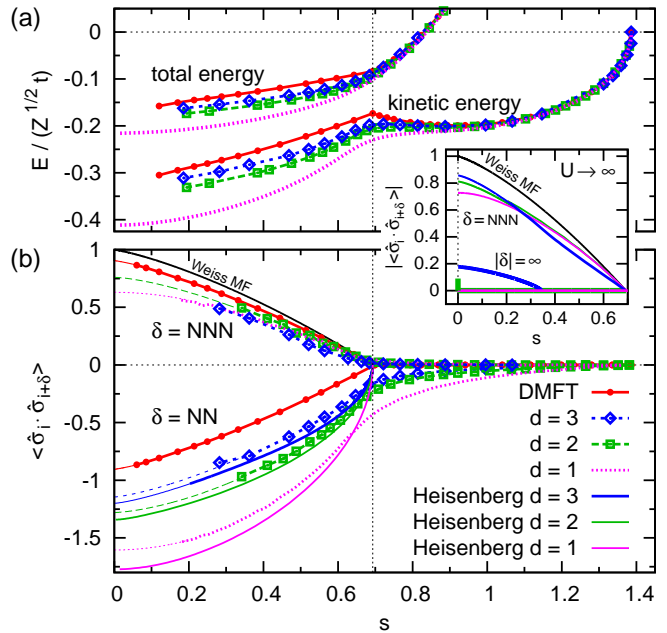}
\caption{(Color online)
Hypercubic lattice at strong coupling: 
(a) Rescaled total and kinetic energy vs entropy.
(b) Spin correlations $\langle\hat{\boldsymbol{\sigma}}_i\cdot\hat{\boldsymbol{\sigma}}_{i+\delta}\rangle$ 
between nearest [$\delta_{d=1}=1$, $\delta_{d=2}=(1,0)$ , $\delta_{d=3}=(1,0,0)$)]
and next-nearest [$\delta_{d=1}=2$, $\delta_{d=2}=(1,1)$, $\delta_{d=3}=(1,1,0)$] neighbors.
Inset: NNN and infinite-range correlations for Heisenberg model ($d=1$ from Ref. \cite{Sato11}).
}\label{fig:dim_ENNvsS}
\end{figure}
%%%%%%%%%%%%%%%%%%%%%%%%%%%%%%%%%%%%%%%%%%%%%%%%%%%%%%%%%%%%%%%%%%%%%%%%%%%%%%%%%

As mentioned in the Introduction, 
nearest-neighbor (NN) spin correlations are, so far, playing a central
role in the experimental quest for AF signatures in cold lattice
fermions. However, as seen in the lower part of
\reff{dim_ENNvsS}(b), the NN spin correlation functions (symbols and
dashed and dotted lines) have strong high-entropy tails in all physical
dimensions $d\le 3$, with only trivial features at $s\approx s^*$,
below which the Heisenberg model (solid lines) becomes applicable.
Conclusions concerning the presence or proximity of AF order could be drawn
from corresponding experimental data only via theoretical look-up
tables.

In contrast, the next-nearest-neighbor (NNN) spin correlation functions
are essentially zero for $s>s^*$ and take off linearly below in all
dimensions. So the mere presence of significant NNN correlations already
implies that the Heisenberg regime $s<\ln(2)$ has been reached.
Remarkably, the results for $d=1$ and $d=2$ are indistinguishable (and
close to DMFT) for $s\gtrsim 0.5$; only those for $d=3$ are slightly
below. The same picture emerges in the Heisenberg model (upper set of
curves in the inset of \reff{dim_ENNvsS}), with slightly larger absolute
values. This near-universality with respect to $d$ establishes that the
crossing point between short-range physics, where spin correlations
increase with lowering $d$, and long-range physics, where spin correlations
decay quickly toward low $d$ (and remain finite in the limit of
distance $\delta\to\infty$ at $s>0$ only for $d\ge 3$ \cite{fn:Heisenberg3d},
cf.\ the lower set of curves in the inset of \reff{dim_ENNvsS}) is
essentially at the NNN distance. 

In this sense, spin correlations at or beyond the NNN distance are much
more representative of genuine AF physics than the unspecific NN
correlations. This is true even regarding LRO, as illustrated in the
Heisenberg limit (chosen due to the higher achievable accuracy) in
\reff{deriv_NN_NNN}:
%%%%%%%%%%%%%%%%%%%%%%%%%%%%%%%%%%%%%%%%%%%%%%%%%%%%%%%%%%%%%%%%%%%%%%%%%%%%%%%%%
\begin{figure}[t]
\includegraphics[width=\columnwidth]{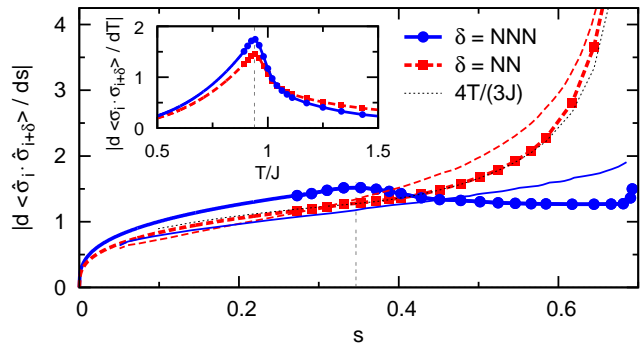}
\caption{(Color online)
Derivatives of NN (squares) and NNN (circles) spin correlation functions of the Heisenberg model 
with respect to 
entropy (main panel) and temperature (inset) in $d=3$. Thin solid and dashed lines: Heisenberg results in $d=2$.
}\label{fig:deriv_NN_NNN}
\end{figure}
%%%%%%%%%%%%%%%%%%%%%%%%%%%%%%%%%%%%%%%%%%%%%%%%%%%%%%%%%%%%%%%%%%%%%%%%%%%%%%%%%
While the entropy derivatives of the NN spin correlations in $d=3$
(squares) are featureless [as those of the spin correlations in $d=2$
(thin lines)], a distinct peak at the critical entropy $s_{\text{N}}$
for AF LRO in $d=3$ \cite{Wessel10} is visible in the NNN data
(circles), with a nearly flat plateau above.  As a comparison with
corresponding $T$ derivatives (with clear peaks at $T_N$ both in the NN
and NNN data -- see the inset of \reff{deriv_NN_NNN}) shows, this qualitative
difference is associated with the use of the entropy as a control
parameter.  With experiments being confined to this entropy
parametrization \cite{fn:entropy2}, it is clear that a detection of the
\neel\ transition via spin correlations would require measurements at
least at NNN distances.

%%%%%%%%%%%%%%%%%%%%%%%%%%%%%%%%%%%%%%%%%%%%%%%%%%%%%%%%%%%%%%%%%%%%%%%%%%%%%%%%%
% Summary
%%%%%%%%%%%%%%%%%%%%%%%%%%%%%%%%%%%%%%%%%%%%%%%%%%%%%%%%%%%%%%%%%%%%%%%%%%%%%%%%%

\myparagraph{Conclusion.}
%In this Letter, we 
We have disentangled generic aspects of
antiferromagnetism both from those specific to infinite-range order
(which is not attainable, by definition, in finite atomic clouds) and
from trivial nearest-neighbor correlations that persist even at high
temperatures, where spin models are not adequate. Our results establish
that the regime $s\lesssim s^* \approx\ln(2)$ is characterized by ``finite-range
antiferromagnetism'' with remarkably universal properties. It may be
detected experimentally, at strong coupling, by a negative slope in
$D(s)$ [or $D(T)$] or by the onset of longer-range (beyond NN) spin
correlations \cite{Pedersen11}. Long-range order appears inessential in
the cold-atom context  and largely decoupled from the basic correlation
mechanisms (cf.\ recent iron layer experiments \cite{Pickel10}):
The mean-field critical entropy $s^*$ is experimentally more relevant than the critical
entropy of an infinite system (in $d=3$) \cite{fn:dynamic}. 
Thus, reaching $s<s_{\text{N}}\approx 0.4$ in cubic systems would not
guarantee additional insight and global equilibration 
(the time scales for which may exceed experimental life times \cite{Schneider12}) is not essential.
Tuning the dimensionality
(and/or adding frustration) for discriminating AF effects appears much more
promising.

We thank M.\ Inoue for help with the BA code, and P.G.J.\ van Dongen,
U.\ Schneider, R.\ P.\ Singh, and L.\ Tarruell for valuable discussions.
Support under ARO Award W911NF0710576 with funds from the DARPA OLE
Program, by CNPq, FAPERJ, and INCT on Quantum Information, and by 
the DFG through SFB/TRR 49 and FOR 1346 is
gratefully acknowledged.

%%%%%%%%%%%%%%%%%%%%%%%%%%%%%%%%%%%%%%%%%%%%%%%%%%%%%%%%%%%%%%%%%%%%%%%%%%%%%%%%%
% Bibliography
%%%%%%%%%%%%%%%%%%%%%%%%%%%%%%%%%%%%%%%%%%%%%%%%%%%%%%%%%%%%%%%%%%%%%%%%%%%%%%%%%


\begin{thebibliography}{16}
\bibitem{fn:length}
Current experimental trap geometries ($\sim 10^5$ fermions) imply AF cores smaller than $40\times 40\times 20$ lattice sites, i.e. each AF site will ``feel'' a boundary within 10 sites.
\bibitem{fn:errors} Trotter errors have been eliminated from the DQMC results and from the Hirsch-Fye QMC \cite{Hirsch86,Bluemer07} DMFT data; finite-size effects are negligible for $U/t=15$ \cite{follow-up}.
\bibitem{fn:scale}
All scales are set by the root mean square energy $\langle \epsilon^2\rangle_{U=0}^{1/2}=\sqrt{Z} t$ (for coordination number $Z$) \cite{Gorelik11}.
\bibitem{fn:Heisenberg3d} Our fit for $d=3$, $\delta=\infty$ is based on the $T=0$ order parameter \cite{Soukoulis91}, $s_N$ \cite{Wessel10}, and the critical exponents.
\bibitem{fn:dynamic}
This is also to be expected for AF signatures in dynamic properties 
such as optical conductivity \cite{Taranto12}.
\end{thebibliography}
\end{document}